\documentclass[epj]{svjour}

\usepackage{graphicx}
\usepackage{fancyhdr}
\usepackage{amssymb}
\def\makeheadbox{{
 }}
\def\mytitle{My title} 
\def\myauthors{My name}  
\def\mytype{My type of session}
\def\mysession{My session}

\def\mytitle{EW NLO corrections to pair production of top-squarks at the LHC}
\def\myauthors{Monika Koll\'ar}    
\def\mytype{Contributed Talk}
\def\mysession{Colliders - SUSY Phenomenology}

\newcommand{\qq}{q\overline{q}} 
\newcommand{\qqbar}{q\overline{q}}

\def \gy{g\gamma}
 
\renewcommand{\Re}{\mbox{Re}}
\def \pt{p_T}
\def \tb{\tan\beta}

\def \msu{m_{\tilde{U}_3}}
\def \msq{m_{\tilde{Q}_3}}

\newcommand{\Oa}{{\cal O}(\alpha)} 
\newcommand{\Oas}{{\cal O}(\alpha^{\mathrm 2})} 
\newcommand{\Oaass}{{\cal O}(\alpha\alpha_{\mathrm s}^{\mathrm 2})}

\newcommand{\Oasc}{{\cal O}(\alpha_{\mathrm s}^{\mathrm 3})} 
\newcommand{\Oasf}{{\cal O}(\alpha_{\mathrm s}^{\mathrm 4})}

\def \mstop{m_{\tilde{t}_1}}

\begin{document}
\title{EW NLO corrections to pair production of top-squarks \\ at the LHC}

\author{Wolfgang Hollik\inst{1},
 Monika Koll\'ar\inst{1} \thanks{\emph{Email:} Monika.Kollar@mppmu.mpg.de}
 \and  Maike K.~Trenkel\inst{1}
}

\institute{Max-Planck Institute for Physics, Munich, Germany}
%
\date{Received: date / Revised version: date}
\date{}
\abstract{
Presented are complete electroweak (EW) corrections at $\Oaass$
to top-squark pair production at the Large Hadron Collider (LHC) within 
the framework of the Minimal Supersymmetric Standard Model (MSSM).
At this order, effects from the interference of EW and QCD contributions  
have to be taken into account. Also photon-induced top-squark production is 
considered as additionnal partonic channel which arises from the non-zero 
photon density in the proton. 
Furthermore, the impact of MSSM parameters on the EW corrections is analyzed.
\PACS{
      {12.15.Lk}{Electroweak radiative corrections}   \and
      {13.87.Ce}{Production}
     } 
} 
\maketitle
\section{Introduction}
\label{intro}

The lighter top-squark is 
as a candidate for the lightest squark within many supersymmetric models
\cite{Ellis:1983ed}, for mainly two reasons based on
the large top 
Yukawa coupling.
Evolving the scalar masses from 
the GUT scale to low scales leads to a low value of the 
top-squark mass, 
and, moreover, a large mixing in the top-squark sector induces a substantial 
splitting between the two mass eigenstates \cite{Djouadi:1996pj}.
Top squarks are therefore of particluar interest, especially 
for hadron colliders.

In hadronic collisions,
top-squarks are primarily produced in pairs via the strong interaction.
Present experimental limits from the Tevatron RUN II data, dependent on the
lightest neutralino mass, are provided by the CDF and D\O~collaborations 
\cite{Aaltonen:2007sw}.
Concerning the theoretical predictions, Born-level cross sections calculated in 
\cite{Kane:1982hw},
have been improved by including the next-to-leading (NLO) corrections in 
supersymmetric QCD (SUSY-QCD). These were worked out in 
\cite{Beenakker:1994an} with the restriction to final state 
squarks of the first two generations. The analysis for the stop sector,
performed in \cite{Beenakker:1997ut}, shows that the SUSY-QCD corrections
significantly modify the LO cross section.

At lowest order in QCD as well as at $\Oasc$, only diagonal top-squarks pairs 
can be produced. 
The non-diagonal production is suppressed as the cross section 
becomes non-zero only at $\Oasf$. The production of non-diagonal top-squark
pairs can also proceed at $\Oas$ via $Z$-exchange in $e^+e^-$ annihilation
\cite{Bartl:1997yi}, or $\qqbar$ annihilation \cite{Bozzi:2005sy}.
The LO cross section for the diagonal pair production depends only
on the mass of the produced squarks.
As a consequence, bounds on the cross section can easily be translated into 
lower bounds on the lightest top-squark mass. At NLO, the cross section
is not only considerably changed, but also other supersymmetric parameters, 
like mixing angle, gluino mass and other sparticle masses, enter
through higher order corrections.
On the other hand, once the top-squarks are discovered, their masses could 
be directly determined from the cross section measurement.

In the following, we study the NLO EW-like corrections to the top-squark pair
production within the Minimal Supersymmetric Standard Model (MSSM). We assume
the MSSM with real parameters, {\em R-parity} conservation and minimal flavor 
violation.

\section{EW NLO Contributions}
\label{sec:class}

As a subset of the
complete set of EW virtual corrections, photonic contributions are present. 
These contain IR singularities, which cancel when also the real photonic 
corrections are taken into account. In addition, at NLO a photon-induced 
subclass 
of contributions appears as an independent production channel.

\subsection{Virtual Corrections}
\label{sec:virt} 

The virtual corrections can be classified according to self-energy, vertex, 
box, and counter-term contributions dressing 
the Born-level partonic amplitudes of 
$\qqbar$ annihilation and gluon fusion. 
Getting an UV finite result requires renormalization 
of the involved quarks and top-squarks.
The counterterms for self-energies, quark and squark vertices and 
the squark quartic interaction at one-loop order are determined
in the on-shell renormalization scheme.
It is not necessary to renormalize the gluon field and the strong coupling
constant.

Loop diagrams involving virtual photons generate IR singularities.
According to Bloch-Nordsieck \cite{Bloch:1937pw}, IR-singular terms cancel
against their counterparts in the real photon corrections.
To regularize the IR singularities we introduce a fictitious photon mass
$\lambda$.
In case of external light quarks, also collinear singularities occur if
a photon is radiated off a massless quark in the collinear limit.
We therefore keep non-zero initial-state quark masses $m_q$ in the loop
integrals, which give rise to single and double logarithmic contributions
of quark masses. The double logarithms cancel in the sum of virtual and real
corrections, single logarithms, however, survive and have to be treated by
means of the factorization.

An additionnal source of IR singularities originates from the gluonic 
insertions to the box contributions in the $\qqbar$ channel. These appear 
in combination with photons or $Z$-bosons.
The similiarity between gluon and photon in the
box contributions allows us to treat the gluonic IR singularities in 
analogy to the photon case.

\subsection{Real Corrections}
\label{sec:real}

\begin{figure}
\includegraphics[width=0.2\textwidth,height=0.13\textwidth,angle=0]{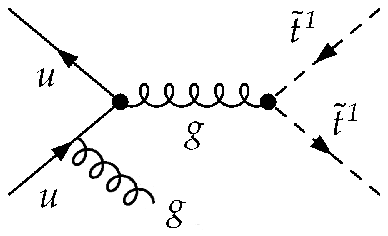} \quad
\includegraphics[width=0.2\textwidth,height=0.15\textwidth,angle=0]{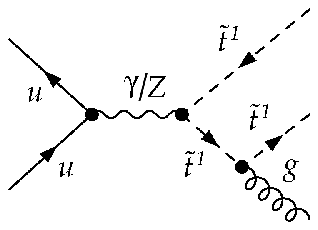}
\caption{Feynman diagrams for gluon bremsstrahlung contributions (left)
   that have to be combined with EW Born diagrams (right). Only interference
   between initial and final state gluon radiation is non-vanishing.}
\label{fig:interf}
\end{figure}

To compensate IR singularities in the virtual EW corrections, real 
photonic and gluonic contributions are required. 
In case of $gg$ fusion, only 
photon brems\-strahlung is needed, whereas in the $\qqbar$ annihilation 
channel, also gluon bremsstrahlung of the appropriate order of $\Oaass$,
shown in Fig.~\ref{fig:interf}, has to be taken into account. 
The necessary contributions originate from the 
interference of QCD and EW tree level diagrams which vanishes at LO.
Not all of the interference terms contribute, however. 
Owing to the color structure, 
only the interference between initial and final state gluon radiation is 
non-zero.

Including the EW--QCD interference in the real corrections does not yet lead 
to an IR-finite result. Also the IR-singular QCD-mediated box corrections 
interfering with the $\Oa$ photon and $Z$-boson tree-level diagrams are 
needed. 
Besides the gluonic corrections there are also IR-finite QCD-mediated 
box corrections, which contain gluinos in the loop. Interfered with the $\Oa$ 
tree-level diagrams, these also give contributions of the respective order 
of $\Oaass$. 

%

We encounter also IR-finite bremsstrahlung contributions, which are, however,
suppressed by more than a factor of ten and therefore not included in our 
numerical studies. For similar reasons we also neglect contributions coming 
from the interference of IR-singular and IR-finite terms.

The treatment of IR-singular brems\-strahlung is done using the phase space 
slicing method. The photonic (gluonic) phase space is split into soft and 
collinear parts, which contain singularities and into non-collinear hard part, 
which is free of singularities and can be integrated numerically.
In the singular regions, the squared matrix elements for the radiative
process factorize into lowest-order matrix elements and universal factors
containig the singularities.

The soft-photon part of the radiative cross section 
in the $\qqbar$ annihilation channel, which is similar to that
in $e^+e^-\rightarrow t\bar t$~\cite{Beenakker:1991ca},
\begin{eqnarray}\label{eq_sigma_soft_phot}
   d\hat{\sigma}_{soft,\gamma}^{q\bar{q}} (\hat s) & = & \frac{\alpha}{\pi}
	\,\Big( e_q^2\,\delta_{soft}^{in} + e_t^2\,\delta_{soft}^{fin}
      + 2 e_q e_t \, \delta_{soft}^{int} \Big) \nonumber \\[1.5ex]
        & \times & \,d\hat{\sigma}_0^{q\bar{q}} (\hat s)\,,
\end{eqnarray}
and for the radiative cross section in the $gg$ fusion channel,
\begin{eqnarray}\label{eq_soft_phOT_fact}
   d\hat{\sigma}_{soft,\gamma}^{gg} (\hat s) & = \frac{\alpha}{\pi}\,
        e_t^2\,\delta_{soft}^{fin} \,d\hat{\sigma}_0^{gg} (\hat s) \,,
\end{eqnarray}
can be expressed using universal factors, $\delta_{soft}^{in, fin, int}$,
which refer to initial state radiation, final state radiation, or
interference of initial and final state radiation, respectively,
with $e_q$ and $e_t$ denoting the electric charges of the initial-state 
quark and of the top-squark, respectively.
$d\hat{\sigma}_0^{q\bar{q} ,gg}$ 
denote the corresponding partonic lowest order cross sections.

The soft-gluon part for the $\qqbar$ channel can be written in a way
similar to (\ref{eq_sigma_soft_phot}), 
but with a different arrangement of the color matrices,
\begin{eqnarray}
   d\hat{\sigma}_{soft,g}^{q\bar{q}} (\hat s) & = & \frac{\alpha_s}{\pi}\,
        \delta_{soft}^{int} \,
      \Big[T^a_{ij} T^a_{lm} T^b_{ji} T^b_{ml}\Big]\,
\\
       &\times& 2 \Re \,\mathrm{\overline{\Sigma}}
		 \left(\widetilde{\mathcal{M}}_{0,g}^{q\bar{q} \, *}
             \widetilde{\mathcal{M}}_{0,\gamma/Z}^{q\bar{q}}\right)\,
      \frac{d\hat{t}}{16\pi \hat{s}^2} \,, \nonumber
\label{eq_sigma_soft_gluon}
\end{eqnarray}
with $\widetilde{\mathcal{M}}$ denoting the ``Born'' matrix elements 
for $g$, $\gamma$ and $Z$ exchange where the color matrices 
are factorized off.

The collinear part of the $2\rightarrow 3$ cross section is proportional
to the Born cross section of the hard process with reduced momentum of
one of the partons. Assuming that parton~$a$ with momentum $p_a$ radiates off a
photon with $p_{\gamma} = (1-z) p_a$, the parton momentum available for the hard
process is reduced to $z p_a$. Accordingly, the partonic energy of the total
process inclusive photon radiation
   $\tilde{\hat{s}} = (p_a + p_b)^2  \,$,
is reduced for the hard process to 
   $\hat{s} = (z p_a + p_b)^2 \,$.

Using these variables, the partonic cross section in the
collinear cones can be written as \cite{Baur:1998kt,Dittmaier:2001ay}
\begin{eqnarray}
   d\hat{\sigma}_{coll} (\hat s) & =&\frac{\alpha}{\pi}\,e_q^2\,
       \int_0^{1-\delta_s}\!\! dz \,\,
       d\hat{\sigma}_0(\hat{s})\,\,\kappa_{coll}(z) \quad , \nonumber\\
\mathrm{with} \qquad
   \kappa_{coll}(z) & = &\frac{1}{2} P_{qq}(z) \biggl[
      \ln \biggl( \frac{\tilde{\hat{s}}}{m_q^2} \, \frac{\delta_{\theta}}{2} \biggr)
      -1 \biggr]
\nonumber \\
		&+& \frac{1}{2} (1-z),
\label{eq_sigma_coll_part}
\end{eqnarray}
where $P_{qq}(z) = (1+z^2)/(1-z)$ is an Altarelli-Parisi splitting function
\cite{Altarelli:1977zs} and $\delta_{\theta}$ is the cut-off parameter to define
the collinear region by $\cos\theta > 1-\delta_{\theta}$.
The Born cross section refers to the hard scale $\hat{s}$,
whereas in the collinear factor the total energy $\tilde{\hat{s}}$ is the
scale needed. In order to avoid an overlap with the soft region, the upper limit
on the $z$-integration in (\ref{eq_sigma_coll_part}) is reduced from $z=1$
to $z=1-\delta_s$, where $\delta_s = 2\Delta E/\sqrt{\hat s}$.

As already mentioned, after adding virtual and real corrections, the mass 
singularity in (\ref{eq_sigma_coll_part}) does not cancel and has to be
absorbed into the (anti-)quark density functions (PDFs). This can be formally 
achieved by a redefinition at NLO QED as shown in 
\cite{Baur:1998kt,Wackeroth:1996hz,Diener:2003ss}.

\subsection{Photon-Induced Top-Squark Pair Production}
\label{sec:phot}

We also consider the photon-induced mechanism of top-squark pair production, which
becomes non-zero at NLO in QED as a direct consequence of the non-zero photon 
density in the proton. Although the photon-induced processes are of different 
overall order, they contribute to the same hadronic final state and thus represent
contributions at NLO QED. We consider only the photon--gluon process in our 
numerical ana\-ly\-sis and neglect the quark--photon process, which gives
contribution of higher order.

As the PDFs at NLO QED have become available only recently \cite{Martin:2004dh},
the results shown here thus correspond to the first study of these effects 
on the top-squark pair production \cite{hollik:07}.

\section{Numerical results}
\label{sec:res}

Here we present
numerical results for the production of lighter top-squark pairs
at LHC energies with EW contributions at one-loop level. We show
in the integrated hadronic cross section $\sigma$ and the differential 
hadronic cross sections with respect to the invariant mass of the top-squark pair 
inclusive the photon ($d\sigma/dM_{inv}$) and to the transverse momentum 
($d\sigma/dp_T$) of one of the final state top-squarks.
All hadronic quantities are obtained by folding the partonic cross section with 
parton distributions and summing over all contributing partons.

Our Standard Model input parameters are chosen in correspondance with 
\cite{AguilarSaavedra:2005pw}. As described above, we use the 
MRST\,2004\,QED~\cite{Martin:2004dh} PDF set with the choice of factorization 
and renormalization scales to equal the sum of the final state particles, 
$\mu_F = \mu_R = 2 m_{\tilde{t}_1}$. $\qq$ denotes the sum of the 
$u\bar{u}$, $d\bar{d}$, $c\bar{c}$, and $s\bar{s}$ annihilation channels.

\begin{table}
\caption{Numerical results for the top-squark pair production at the LHC within 
different SPS mSUGRA scenarios 
\protect\cite{Allanach:2002nj,AguilarSaavedra:2005pw}
(the values of $\sigma$ and $\Delta\sigma$ are given in fb).}
\label{tab_results}
\begin{tabular}{ccrr@{}lr@{}l@{}l}
\hline\noalign{\smallskip}
scenario & channel & $\sigma^{LO}$ & 
\multicolumn{2}{c}{$\Delta \sigma^{NLO}$} & 
\multicolumn{3}{l}{$\frac{\Delta \sigma^{NLO}}{\sigma^{LO}}$} \\ 
\noalign{\smallskip}\hline\hline\noalign{\smallskip}
SPS1a & $\qq$      &  220 &  -9&.65 & -4&.4&\% \\[0.5mm]
      &   gg       & 1444 & -15&.4  & -1&.1&\% \\[0.5mm]
      & $\gy$      &      &  29&.0  &   &   &  \\[0.5mm]
      & \bf{total} & 1664 &   3&.95 &  0&.24&\%\\
\noalign{\smallskip}\hline\noalign{\smallskip}
SPS1a'& $\qq$      &  436 & -11&.5  & -2&.6&\% \\[0.5mm]
      &   gg       & 3292 & -14&.6  & -0&.44&\%\\[0.5mm]
      & $\gy$      &      &  58&.5  &   &    & \\[0.5mm]
      & \bf{total} & 3728 &  32&.4  &  0&.87&\%\\
\noalign{\smallskip}\hline\noalign{\smallskip}
SPS2  & $\qq$      & 1.16 & -8.98&$\times 10^{-2}$ & -7&.7&\% \\[0.5mm]
      &   gg       & 2.97 & -3.07&$\times 10^{-2}$ & -1&.0&\% \\[0.5mm]
      & $\gy$      &      &  0.15&5                &   &   &  \\[0.5mm]
      & \bf{total} & 4.13 &  3.45&$\times 10^{-2}$ &  0&.84&\%\\
\noalign{\smallskip}\hline\noalign{\smallskip}
SPS5  & $\qq$      & 2870 &-1&3.2                  &  0&.46&\%\\[0.5mm]
      &   gg       &31960 &49&9                    &  1&.6&\% \\[0.5mm]
      & $\gy$      &      &40&5                    &   &   &  \\[0.5mm]
      & \bf{total} &34830 &89&1                    &  2&.6&\% \\
\noalign{\smallskip}\hline
\end{tabular}
\end{table}

In~Table~\ref{tab_results} we show results for the total hadronic cross sections 
within four different SPS mSUGRA scenarios 
\cite{Allanach:2002nj,AguilarSaavedra:2005pw}.
The total cross sections at leading order, $\sigma^{LO}$,
the absolute size of the EW corrections, which corresponds to the difference 
between the LO and NLO cross sections, $\Delta\sigma^{NLO}$, and the relative 
corrections, $\delta$, given as the ratio of NLO corrections to the respective LO 
contributions, are presented for  $gg$ fusion, $\qq$ annihilation, and $\gy$ 
fusion separately. The $\gy$ channel contributes only at NLO. The total amount of 
EW corrections is small in all SUSY scenarios.

\begin{figure}
\includegraphics[width=0.45\textwidth,height=0.57\textwidth,angle=0]{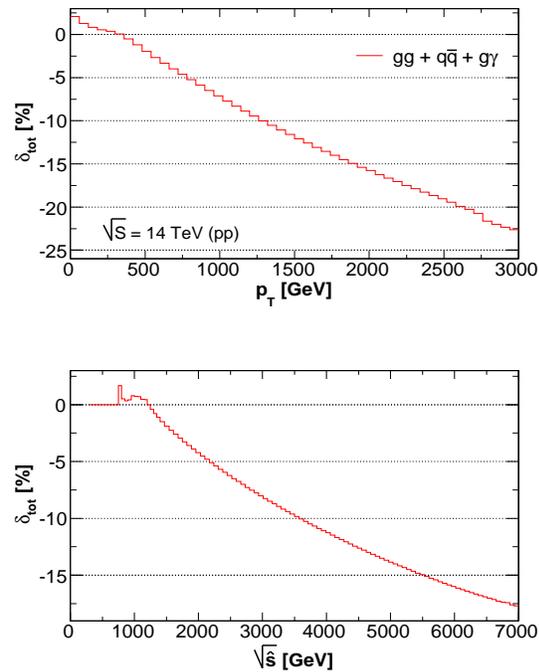}
\caption{Overall relative corrections 
$\delta_{tot}$ with respect to $\pt$ and $\sqrt{\hat s}$ at the LHC
within the SPS1a scenario. }
\label{fig:relcorr}
\end{figure}

In order to illustrate the numerical impact of the EW corrections on the LO 
cross section, in Fig.~\ref{fig:relcorr} we show the overall relative corrections 
$\delta_{tot} = \Delta \sigma^{NLO}/\sigma^{LO}$ as distributions with respect to 
$\pt$ and $\sqrt{\hat s}$ (which corresponds to $M_{inv}$). 
In the $\pt$-distribution, the corrections grow in size with increasing $\pt$ and
reach about $-20\%$ for $ \pt \gtrsim 2500$~GeV. Similar effects are visible in 
the $M_{inv}$ distribution. Although smaller in size, the corrections raise up to
$-15\%$ level for $\sqrt{\hat s} \gtrsim 5000$~GeV. These ranges of $\pt$ and 
$\sqrt{\hat s}$ are still within the reach of the LHC.

The behavior of EW corrections at high scales is dominated by the massive gauge 
boson contributions, which consist of double logarithms of $W$ and $Z$ masses. 
The double logarithms are not canceled by the real gauge boson corrections,
since these correspond to different hadronic final states. As a result, 
large negative contributions show up in the $\pt$ and $\sqrt{\hat s}$
distributions.

It is obvious that although small for the total cross sections, the EW 
corrections cannot be neglected in the differential distributions, since 
in the high-$\pt$ and high-$\sqrt{\hat s}$ ranges they are of the same order 
of magnitude as the SUSY-QCD corrections.

\begin{figure}
\includegraphics[width=0.45\textwidth,height=0.28\textwidth,angle=0]{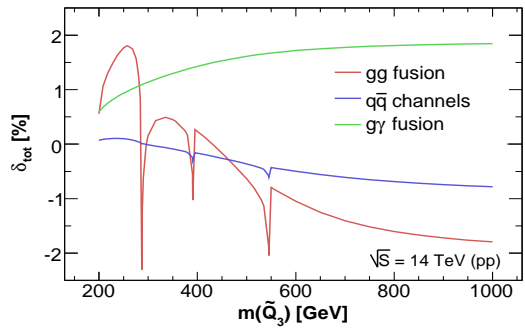}
\caption{ Relative EW corrections as a function of the soft-breaking
   parameter $\msq$ for each of the indicated channels compared to the total 
	($gg + \qq$) LO cross section at the LHC, all other parameters are fixed 
	within the SPS~1a' scenario.}
\label{fig:msq_delta}
\end{figure}

We have also studied the dependence of the EW contributions on various SUSY 
parameters. We have focused on the parameters that affect the top-squark mass, 
since here the effects are expected to be strongest. Following parameters have been
varied: $\msq$, $\msu$, $\tb$, $A_t$, and $\mu$ around the SPS~1a' value, while 
keeping all other parameters fixed. As an example, in Fig.~\ref{fig:msq_delta} 
we show the dependence of the overall EW corrections on 
$\msq$ for each production channel separately.

The $\gy$ corrections grow up to $2\%$ with increasing $\mstop$ in the 
considered range. The $\gy$ fusion channel is as important as the $\qq$ and 
$gg$ channels and should be taken into account for a reliable cross section 
prediction. The $\qq$ corrections involve many different SUSY particles in the 
loops, however the relative corrections change only little between different
scenarios, varying between $0\%$ and $-1\%$. 
The $gg$ contributions are more sensible to variations of the considered SUSY 
parameters. The general behavior is similar to the $\qq$ case, the 
decrease with increasing $\mstop$ is, however, stronger. The $gg$ plot is 
dominated by striking (negative) peaks, some of them are 
(although much weaker) also visible in the $\qq$ corrections.

The peaks originate from two sources of threshold effects. One of them is the 
Higgs boson $H^0$ threshold, $m(H^0) = 2\mstop$, which affects only the $gg$ 
channel. Second source of the threshold effects
enhances the corrections in scenarios where $\mstop$ 
equals the sum of masses of a neutralino and the top-quark or of a chargino 
and the bottom-quark. 
The parameter regions with large EW corrections due to the threshold 
effects are small and over the wide range of SUSY parameters, the EW 
corrections to the top-squark pair production are smaller than $1\%$.

\begin{thebibliography}{999}
%
%
\bibitem{Ellis:1983ed}
J.~R. Ellis and S.~Rudaz, {\em Phys. Lett.} {\bf B128} (1983) 248.
\bibitem{Djouadi:1996pj}
A.~Djouadi, J.~Kalinowski, P.~Ohmann, and P.~M. Zerwas, 
  {\em Z. Phys.} {\bf C74} (1997)
  93--111, {\tt hep-ph/9605339}.
\bibitem{Aaltonen:2007sw}
{\bf CDF} Collaboration, T.~Aaltonen, {\tt 0707.2567} \\
{\bf D0} Collaboration, V.~M. Abazov {\em et.~al.},
	{\em Phys. Lett.} {\bf B645} (2007) 119--127, 
	{\tt   hep-ex/0611003}.
\bibitem{Kane:1982hw}
G.~L. Kane and J.~P. Leveille, {\em Phys. Lett.} {\bf B112} (1982) 227 \\
P.~R. Harrison and C.~H. Llewellyn~Smith, {\em Nucl. Phys.} {\bf B213} 
	(1983) 223 \\
E.~Reya and D.~P. Roy, {\em Phys. Rev.} {\bf D32} (1985) 645 \\
S.~Dawson, E.~Eichten, and C.~Quigg, {\em Phys. Rev.} {\bf D31} 
	(1985) 1581 \\
H.~Baer and X.~Tata, {\em Phys. Lett.} {\bf B160} (1985) 159.
\bibitem{Beenakker:1994an}
W.~Beenakker, R.~H$\ddot{\mathrm o}$pker, M.~Spira, and P.~M. Zerwas, 
  {\em Phys. Rev. Lett.} {\bf 74} (1995) 2905--2908, 
  {\tt hep-ph/9412272}, {\em Nucl. Phys.} {\bf B492} (1997) 51--103, 
  {\tt   hep-ph/9610490}.
\bibitem{Beenakker:1997ut}
W.~Beenakker, M.~Kr$\ddot{\mathrm a}$mer, T.~Plehn, M.~Spira, and P.~M. Zerwas,
  {\em Nucl. Phys.} {\bf B515} (1998) 3--14, 
  {\tt hep-ph/9710451}.
\bibitem{Bartl:1997yi}
A.~Bartl {\em et.~al.}, {\em Z. Phys.} {\bf C76} (1997) 549--560,
  {\tt hep-ph/9701336}.
\bibitem{Bozzi:2005sy}
G.~Bozzi, B.~Fuks, and M.~Klasen, {\em Phys. Rev.} {\bf D72} (2005) 035016,
  {\tt hep-ph/0507073}.
\bibitem{Bloch:1937pw}
F.~Bloch and A.~Nordsieck, {\em Phys. Rev.} {\bf 52} (1937) 54--59.
\bibitem{Beenakker:1991ca}
W.~Beenakker, S.~C. van~der Marck, and W.~Hollik, {\em Nucl. Phys.} {\bf
  B365} (1991) 24--78.
\bibitem{Baur:1998kt}
U.~Baur, S.~Keller, and D.~Wackeroth, {\em Phys. Rev.} {\bf D59}
  (1999) 013002, {\tt hep-ph/9807417}.
\bibitem{Dittmaier:2001ay}
S.~Dittmaier and M.~Kr$\ddot{\mathrm a}$mer, {\em Phys. Rev.}
  {\bf D65} (2002) 073007, {\tt hep-ph/0109062}.
\bibitem{Altarelli:1977zs}
G.~Altarelli and G.~Parisi, {\em Nucl. Phys.} {\bf B126} (1977) 298.
\bibitem{Wackeroth:1996hz}
D.~Wackeroth and W.~Hollik, {\em Phys. Rev.} {\bf D55} (1997)
  6788--6818, {\tt hep-ph/9606398}.
\bibitem{Diener:2003ss}
K.~P.~O. Diener, S.~Dittmaier, and W.~Hollik, {\em Phys. Rev.} {\bf D69} 
	(2004) 073005, {\tt hep-ph/0310364}.
\bibitem{Martin:2004dh}
A.~D. Martin, R.~G. Roberts, W.~J. Stirling, and R.~S. Thorne,
  {\em Eur. Phys. J.} {\bf C39} (2005) 155--161, 
  {\tt hep-ph/0411040}.
\bibitem{hollik:07}
W.~Hollik, M.~Kollar and M.~K.~Trenkel {\em MPP-2007-149}, to be published.
\bibitem{Allanach:2002nj}
B.~C. Allanach {\em et.~al.}, {\tt hep-ph/0202233}.
\bibitem{AguilarSaavedra:2005pw}
J.~A. Aguilar-Saavedra {\em et.~al.},
  {\em Eur. Phys. J.} {\bf C46} (2006) 43--60,
  {\tt hep-ph/0511344}.


\end{thebibliography}
%

\end{document}